\documentclass[aps,twocolumn,superscriptaddress]{revtex4}
\usepackage{amsmath}
\usepackage{color}
\usepackage{bbm}
\usepackage{amssymb}
\usepackage{epsfig}
\usepackage{multirow}
\usepackage{amsbsy}
\usepackage{array}
\usepackage{diagbox}
\usepackage{bm}
\usepackage{tikz}
\usepackage{extarrows}
\usepackage{graphicx}
\usepackage{appendix}
\usepackage{txfonts}
\usepackage[normalem]{ulem}
\usepackage[colorlinks=true,linkcolor=blue,citecolor=blue,urlcolor=blue,bookmarks=false]{hyperref}

\definecolor{orange}{rgb}{1,0.5,0}

\begin{document}
\title{Floquet Prethermal Order by Disorder}
\date{\today}
\author{Hui-Ke Jin}
\affiliation{Technical University of Munich, TUM School of Natural Sciences, Physics Department, 85748 Garching, Germany}
\affiliation{Munich Center for Quantum Science and Technology (MCQST), Schellingstr. 4, 80799 M{\"u}nchen, Germany}

\author{Johannes Knolle}
\affiliation{Technical University of Munich, TUM School of Natural Sciences, Physics Department, 85748 Garching, Germany}
\affiliation{Munich Center for Quantum Science and Technology (MCQST), Schellingstr. 4, 80799 M{\"u}nchen, Germany}
\affiliation{Blackett Laboratory, Imperial College London, London SW7 2AZ, United Kingdom}

\begin{abstract}
Frustrated magnets can have accidental ground state degeneracies which may be lifted by various forms of disorder, for example in the form of thermal or quantum fluctuations. This {\it order by disorder} (ObD) paradigm is well established in equilibrium and here is generalized to Floquet many-body systems. Investigating a periodically-driven XXZ-compass model on the square lattice, we show that in a prethermal regime, dynamical fluctuations induced by high-frequency drives select a discrete set of states out of a degenerate ground state manifold of the lowest order Floquet Hamiltonian chosen as initial states. Remarkably, prior to the ObD selection, an unusual fluctuating regime emerges leading to a prethermalization timescale scaling linearly with the drive frequency. We argue that prethermal ObD with its unusual approach to the selected states is a generic phenomenon of driven frustrated systems and confirm it in the paradigmatic $J_1-J_2$ XX model.
\end{abstract}
\maketitle

{\em Introduction.---}
Frustration in many-body systems originates from the presence of competing interactions~\cite{Vannimenus1977,Toulouse1987,BookDiep,BookLacroix} and can lead to exotic phenomena such as classical spin liquids, e.g., spin ices~\cite{spinice,castelnovo2008magnetic}, or topologically-ordered ground states in quantum spin liquids~\cite{Lee08,Balents10,QSLRMP,Broholm2020,knolle2019field,savary2016quantum}.  Typically, a system subject to frustration can exhibit {\em accidental} degeneracies of the classical ground-state (g.s.) manifold, unprotected by symmetry. A central concept in the field is the {\em order by disorder} (ObD) mechanism~\cite{orderbyd1,orderbyd2,shender1996order,chalker2017spin}, which entails that the classical degeneracy can be lifted by introducing fluctuations (or entropy). Although the latter normally favor featureless disordered states, via ObD they turn out to {\em induce} long-range ordering. 

In equilibrium there are typically three distinct forms of ObD: i) Originally Villain and collaborators showed that disorder in the form of site dilution can induce long-range order in a fully frustrated Ising model~\cite{orderbyd1}. Generically, quenched disorders can induce a discrete set of minima in an otherwise degenerate manifold of the free energy~\cite{prakash1990ordering}. ii) Disorder as thermal fluctuations can lead to an entropic selection of specific states from the degenerate manifold, eventually restoring long-range order at finite temperatures~\cite{Henley1989,chalker1992hidden,Moessner1998}. iii) Zero point quantum fluctuations can similarly promote ordered states even at zero temperature~\cite{Henley1989,Chubukov1992}.  Up to now, most studies of ObD have been confined to the realm of equilibrium physics. Notable exceptions are works on tuning the free energy landscape via time-dependent modulations~\cite{Wan2017, Wan2018} and, very recently, the generalization of ObD to dynamical systems with non-reciprocal interactions~\cite{Hanai2024}, both of which incorporate stochastic noise for stabilizing ObD-selected steady states. %{\bf CHECK!} 

In this paper, we address the question of whether ObD can be generalized to driven systems without noise and whether qualitatively new features can arise with time dependence?
We provide affirmative answers by focusing on Floquet frustrated magnets.  Floquet engineering under periodic driving has recently become a versatile tool for realizing novel phases of matter beyond thermal equilibrium~\cite{Bukov2015, Eckardt2017,Oka2019,Rudner2020}.  Even in the absence of external noise, driven systems can settle - after a short typically frequency independent time scale $\tau_{pth}$ -- into quasi-steady, so-called {\it prethermal}, states. These appear for large drive frequency $\omega$ and can stabilize quasi-equilibrium and non-trivial dynamical phases up to exponentially long (in $\omega$) thermalization timescales, after which the system heats up to a featureless infinite temperature state~\cite{Berges2004, Bukov2015PRL, Abanin2015, Canovi2016, Mori2016, Weidinger2017, Abanin2017CMP, Mori2018,Jin2022,Yue2023}. While this has been first worked out for quantum many-body systems~\cite{Abanin2015,Mori2016}, similar effects appear in classical systems where they can be efficiently simulated~\cite{Andrea2021,Ye2021,Andrea2021PRB,Yue2023PRL}.
It allows us to show that the fluctuations induced by a periodic drive can select a long-range ordered prethermal phase, a phenomenon we dub {\em Floquet prethermal ObD}. Remarkably, the transition in time to the prethermal plateau is itself highly unusual. Namely, starting from states close to the degenerate g.s. manifold (of the Floquet Hamiltonian in high-frequency limit), a long fluctuating regime appears scaling linear in frequency, $\tau_{pth}\propto \omega$. The characteristic scaling appears because of a hierarchical symmetry reduction of the effective Floquet Hamiltonian as recently introduced in Ref.~\cite{Fu2024}. Intuitively, the systems need sufficient time for the drive-induced fluctuations to explore the whole phase space to generate the entropy for prethermal ObD selection.    

\begin{figure*}[!htp]
\includegraphics[width=\linewidth]{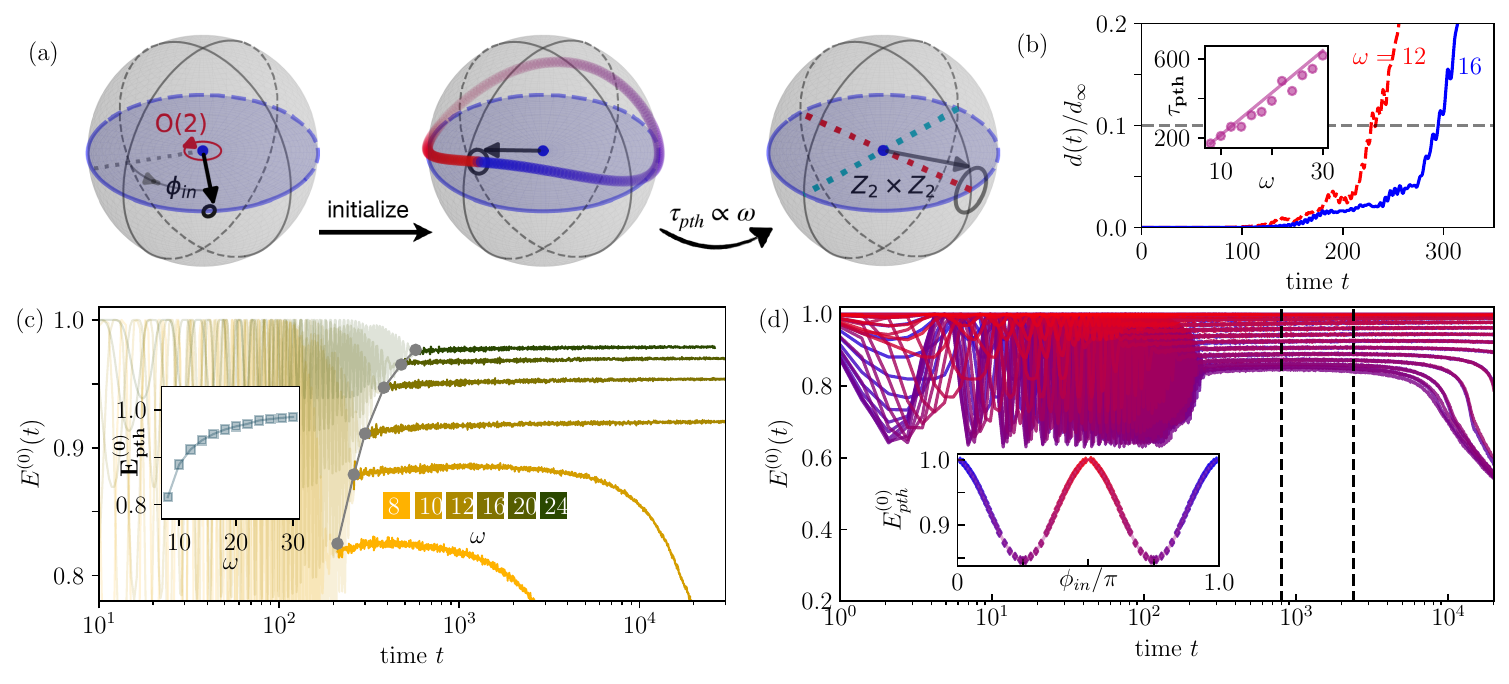}
\caption{Prethermal order-by-disorder. {\bf (a)} The dynamics of an individual O(3) spin (arrow), and the circles positioned atop the arrows strand for the averaged fluctuations. Left: The system has an {\em accidentally} O(2)-degenerate g.s. manifold about the $z$-axis, w.r.t. the zeroth-order effective Hamiltonian $H^{0}_{\rm eff}$ in Eq.~\eqref{eq:H0}. The initial state manifests as an in-plane FM-ordered state with an azimuthal angle $\phi_{in}$. Middle: Before the prethermal phase, all spins evolve as a single spin, tracing a trajectory represented by colorful dots (blue to red over time). The fluctuations (circle) increase as $~e^{t/\tau_{pth}}$. Right: In the prethermal phase, fluctuations are relatively strong and almost stop increasing, leading to a FM-ordered prethermal state selected by fluctuations. {\bf (b)} The decorrelator initially grows exponentially. The cross of the dashed line at 0.1 and $d(t)$ determines the prethermalization timescale $\tau_{pth}$, which is proportional to drive frequency $\omega$, see inset. {\bf (c)} Dynamics of the zero-th order energy $E^{(0)}(t)$. The transparent (solid) regime represents the timescale before (after) $\tau_{pth}$. Inset: The zeroth order prethermal energy $E^{(0)}_{pth}$ as a function of drive frequency $\omega$. {\bf (d)} $E^{(0)}(t)$ for various initial azimuthal angles $\phi_{in}$, varying from $\phi_{in}=0$ to $\phi_{in}=\pi/2$ (blue to red). Note that the blue-side line ($\phi^b_{in}\in[0,\pi/4]$) and corresponding red-side line $\phi^r_{in}=\pi/2-\phi^b_{in}$ converge into a single line in the prethermal regime. Inset: The zeroth-order prethermal energy $E^{(0)}_{pth}$ as a function of $\phi_{in}$. $E^{(0)}_{pth}$ is calculated as an average over the time interval of $[10^3, 3\times{}10^3]$, corresponding to the time window delimited by two vertical dashed lines. The size of the marker is magnified by a factor of $10^3$ times the standard deviation.  Here we used $L=72$, $K=1$, $J^z=0$, and $G=10^{-3}$ [$\phi_{in}=0.2\pi$ for (b, c) and $\omega=9$ for (d)].
}\label{fig:fig1}
\end{figure*}

To illustrate this phenomenon, we consider a classical frustrated compass spin model on the square lattice which displays an accidentally $O(2)$ degenerate g.s. manifold. It was recently shown that in equilibrium, four discrete states are selected via thermal ObD~\cite{Khatua2023}. We here generalize the model to a periodically driven version. We show that a characteristic dependence of the prethermal mean energy and thermalization time appears for initial states chosen from the original $O(2)$ manifold. Independent of the initial state, each trajectory settles into four discrete states and the selection mechanism is always accompanied by a long fluctuating regime. To explain the latter, we develop an analytical theory of harmonic fluctuations around the mean-field transient dynamics. Frequency-dependent corrections to the effective Floquet Hamiltonian lead to an increase of fluctuations in time, thereby accounting for the unusual prethermalization timescale $\tau_{pth}\propto \omega$. 

Finally, we argue that Floquet prethermal ObD is a generic phenomenon originating from the interplay of accidental degeneracies and periodic driving. We confirm that it also appears in the paradigmatic $J_1-J_2$ model on the square lattice and discuss its experimental relevance.

{\em Model.---}
We study a trinary Floquet Hamiltonian $H(t)$ on the square lattice, subjected to a periodic modulation at frequency $\omega=2\pi/T$ (with period $T$),
\begin{equation}
H(t) = 
\begin{cases}
H_{x}\equiv{}-\sum_{r,\delta}J^x_{\delta{}}S^x_{r}S^x_{r+\delta{}} & {\rm~for~} t\in[0,~T/3)\\
H_{y}\equiv{}-\sum_{r,\delta}J^y_{\delta{}}S^y_{r}S^y_{r+\delta{}} & {\rm~for~} t\in[T/3,~2T/3)\\
H_{z}\equiv{}-\sum_{r,\delta}J^z{}S^z_{r}S^z_{r+\delta{}} & {\rm~for~} t\in[2T/3,~T),
\end{cases}
\label{eq:Ht}
\end{equation}
where $r$ is the lattice site and $\delta=\hat{x},\hat{y}$ denote the lattice vector along the $x$- and $y$-directions, respectively. %Specifically, we consider a dynamical XXZ-compass interaction such that $J^a_{\delta} =J+K{}\Theta_{a,\delta}$ ($a=x,y$), where  $\Theta_{x,\hat{x}(y,\hat{y})}=1$ and it takes zero otherwise.
Specifically, the bond-oriented XXZ-compass interaction $J^a_{\delta}$ is graphically illustrated as~\cite{appendix}
$$\begin{tikzpicture} 
\draw (-5.8,0) node {$J^{x}_{\delta}=$} ;
\draw[dotted][very thick][blue] (-5.27, 0.27)--(-4.73, 0.27); 
\draw[dotted][very thick][blue] (-5.27,-0.27)--(-4.73,-0.27); 
\draw[very thick][orange]   (-5.27,-0.27)--(-5.27, 0.27);
\draw[very thick][orange]   (-4.73,-0.27)--(-4.73, 0.27);
\draw (-4.6,-0.1) node {,};

\draw (-5.8+1.9,0) node {$J^{y}_{\delta}=$} ;
\draw[very thick][orange]   (-5.27+1.9, 0.27)--(-4.73+1.9, 0.27); 
\draw[very thick][orange]   (-5.27+1.9,-0.27)--(-4.73+1.9,-0.27); 
\draw[dotted][very thick][blue] (-5.27+1.9,-0.27)--(-5.27+1.9, 0.27);
\draw[dotted][very thick][blue] (-4.73+1.9,-0.27)--(-4.73+1.9, 0.27);
\draw (-4.6+1.9,-0.1) node {,};

\draw (-4.6+2.5,0.) node {for};

%\draw[dotted][very thick]  (-5.27+3.8,-0.25)--(-5.27+3.8, 0.25);
\draw[dotted][very thick][blue]  (-5.27+3.8-0.2,-0.)--(-5.27+3.8+0.2, 0.);
\draw (-4.6+4.0,0) node {$\equiv J+K$};
\draw (-4.6+4.6,-0.1) node {,};

%\draw[very thick][blue]  (-5.27+5.7,-0.25)--(-5.27+5.7, 0.25);
\draw[very thick][orange]  (-5.27+5.8-0.2,-0.)--(-5.27+5.8+0.2, 0.);
\draw (-4.6+5.7,0) node {$\equiv J$};
\draw (-4.6+6.,-0.1) node {.};
\end{tikzpicture}$$
Here, we use $J=1$ as the energy unit and restrict to $K{}\geq{}0$ and $0\leq{}J^z<1$.
The spin dynamics is described by the standard Hamilton equations of motion $\partial\vec{S}_{r}/\partial{}t=\{\vec{S}_r,~H(t)\}.$ It leads to a stroboscopic evolution function which can be analytically integrated~\cite{Howell2019,appendix}. %Hence the stroboscopic equation of motion for a spin is $\vec{S}_r(t+T) = \mathsf{R}_{z}[\theta^z_r(t)]\mathsf{R}_{y}[\theta^y_r(t)]\mathsf{R}_{x}[\theta^x_r(t)]\vec{S}_{r}(t),$ where $\mathsf{R}_{a}[\theta^a_r]$ ($a=x,y,z$) denotes rotation matrices about the $a$ axis by an angle $\theta^a_r= -\frac{T}{3}\sum_{\delta=\pm{}\hat{x},\pm{}\hat{y}}J^a_{\delta{}}S^a_{r+\delta{}}$ (we use the notation of $J^a_{-\delta}\equiv{}J^a_{\delta}$ and $J^z_\delta{}\equiv{}J^z$).
%Here we focus on the s
Noe that the stroboscopic time dependence is chosen for numerical convenience but a continuous drive would lead to similar results. 

Given the stroboscopic evolution function, the effective prethermal Hamiltonian $H_{\rm eff}$ is defined as $e^{-T\{H_{\rm eff},~\cdot\}}\equiv{}e^{-\frac{T}{3}\{H_z,~\cdot\}}e^{-\frac{T}{3}\{H_y,~\cdot\}}e^{-\frac{T}{3}\{H_x,~\cdot\}}.$
In the high-frequency limit, we can expand the above equation %by using Baker-Campbell-Hausdorff formula, 
and approximate the effective Hamiltonian up to the first order as~\cite{Mori2018Classical}
$H_{\rm eff}=H_{\rm eff}^{(0)}+H_{\rm eff}^{(1)}+O(\omega^{-2}).$
The zeroth order, $H_{\rm eff}^{(0)}$, is just the XXZ-compass model studied in Ref.~\cite{Khatua2023}
\begin{equation}
H_{\rm eff}^{(0)}=\left(H_{x}+H_{y}+H_{z}\right)/3.\label{eq:H0}
\end{equation}
The sub-leading  order, which comes from the Poisson brackets $H_{ab}\equiv{}\{H_{a},~H_{b}\}$, can be expressed as 
\begin{equation}
H_{\rm eff}^{(1)}=-\frac{\pi}{9\omega}\left(H_{zx}+H_{zy}+H_{yx}\right),\label{eq:H1}
\end{equation}
where 
$H_{ab}=
\sum_{r,\delta=\pm{}\hat{x},\pm{}\hat{y}}\sum_{\delta'\neq{}\delta}J^{a}_{\delta}J^b_{\delta'}\epsilon^{abc}S^a_{r+\delta}S^b_{r+\delta'}S^c_{r}+\sum_{r,\delta=\hat{x},\hat{y}}J^{a}_{\delta}J^b_{\delta}\epsilon^{abc}\left(S^a_rS^b_rS^c_{r+\delta} + S^a_{r+\delta}S^b_{r+\delta}S^c_{r}\right). $   
%\begin{equation*}
%\begin{split}
%H_{ab}=%&\sum_{r,\delta=\hat{x},\hat{y}}\sum_{\delta'\neq{}\delta}J^{a}_{\delta}J^b_{\delta'}\epsilon^{abc}\left(S^a_{r+\delta}S^b_{r+\delta'}S^c_{r}+S^a_{r+\delta}S^b_{r-\delta'}S^c_{r}\right)+\\
%&\sum_{r,\delta=\pm{}\hat{x},\pm{}\hat{y}}\sum_{\delta'\neq{}\delta}J^{a}_{\delta}J^b_{\delta'}\epsilon^{abc}S^a_{r+\delta}S^b_{r+\delta'}S^c_{r}+\\
%&\sum_{r,\delta=\hat{x},\hat{y}}J^{a}_{\delta}J^b_{\delta}\epsilon^{abc}\left(S^a_rS^b_rS^c_{r+\delta} + S^a_{r+\delta}S^b_{r+\delta}S^c_{r}\right).    
%\end{split}
%\end{equation*}
In the absence of the compass anisotropy, $K=0$, $H^{(0)}_{\rm eff}$ in Eq.~\eqref{eq:H0} exhibits in-plane $O(2)$ symmetry about the $z$-axis. The g.s. is given by ferromagnetic (FM) aligned spins within the XY plane, characterized by an azimuthal angle $\phi\in[0,2\pi)$.
For $K>0$, the $O(2)$ symmetry of $H^{(0)}_{\rm eff}$ is reduced into a discrete $Z_4$ one. However, the g.s. energy remains invariant under arbitrary global in-plane spin rotations, manifesting an {\em accidentally} $O(2)$-degenerate g.s. manifold, see Fig.~\ref{fig:fig1}(a).  In equilibrium, fluctuations of thermal ObD lift the accidental degeneracy and the free energy has four minima at half-integer $\pi$~\cite{Khatua2023,appendix}.  In the driven case, we observe that $H_{\rm eff}^{(1)}$ reduces the $Z_4$ symmetry to $Z_2$, which is an example of the hierarchical symmetry reduction of Floquet systems~\cite{Fu2024}.% {\bf CHECK AND DISCUSS}
%The first frequency dependent correction also energetically favors out-of-plane correlations, which can lead to a dynamical phase transition as we show below.  

We first explore the dynamical model in Eq.~\eqref{eq:Ht} via numerical simulations on a finite system with $N=L^2$ spins. Since $H^{(0)}_{\rm eff}$ exhibits thermal ObD, it is natural to explore whether the corresponding non-equilibrium dynamics can establish a similar state selection mechanism via prethermalization. Therefore, we initialize the simulations with the FM ordered g.s. of $H^{(0)}_{\rm eff}$, characterized by a spin orientation $(\cos\phi_{in}, \sin\phi_{in},0)$, where $\phi_{in}$ denotes the initial azimuthal angle, see Fig.~\ref{fig:fig1}(a). To bring the many-body character of the system into play, we add perturbations on top of this FM order such that $\phi_r=\phi_{in}+\delta{}\theta_r$, with $\delta\phi_r$'s a random Gaussian noise with a standard deviation of $2\pi{}G$. Nonzero $G$ increases the energy density of initial states and tunes the effective temperature of the expected prethermal plateau. 

{\em Prethermalization timescale.---} 
Generic periodically driven many-body systems with large drive frequencies are governed by two timescales. The first one, $\tau_{pth}$, captures the initial equilibration dynamics after which the system enters the prethermal plateau. Normally as the short-time dynamics is primarily governed by $H^{(0)}_{\rm eff}$, the prethermalization time is independent of frequency and governed by the local energy scale, $\tau_{pth}\propto 1/J$.  However, as a key finding, we report that in our set-up,  $\tau_{pth}$ has an unusual dependence on $\omega$ as $\tau_{pth}\propto\omega$, see inset of Fig.~\ref{fig:fig1}(b). To probe $\tau_{pth}$, we introduce the decorrelator  $d(t)$ quantifying the distance between two initially close replicas of the system~\cite{Andrea2021,Andrea2021PRB,Bilitewski2018,Bilitewski2021,appendix}, $\vec{S}_r(r)$ and $\vec{S}'_r(t)$, as 
$d(t)=\sqrt{\frac{1}{N}\sum_{r}\left[\vec{S}_r(t)-\vec{S}'_r(t)\right]^2}.$ During the initial fluctuating dynamics the spins move together and the value of the decorrelator remains small. Eventually, it grows exponentially from its tiny value around $d(t=0)$, and this onset of many-body chaos means that the system explores the many-body phase space (the distribution of initially almost aligned spins spreads). Note, at infinite temperature characterized by only trivial correlations, the typical value of decorrelator is $d_{\infty}=\sqrt{2}$~\cite{Andrea2021}. The prethermalization timescale $\tau_{pth}$ is quantitatively determined by the time at which the decorrelator crosses 10\% of its infinite temperature from a tiny initial value. %see Fig.~\ref{fig:fig1}(b).

Intuitively, $\tau_{pth}$ quantifies the time when the dynamics of systems deviates from its motion as a single large spin, that is, from the mean-field solution $\bar{S}^a(t)$ given by the generalized Landau-Lifshitz-Gilbert (LLG) equation
\begin{equation}
	\partial_t\bar{S}^a(t)	= \{\bar{S}^a(t), H^{(0)}_{\rm eff}\}+\{\bar{S}^a(t), H^{(1)}_{\rm eff}\}+O(\omega^{-2}).\label{eq:absLLG}
\end{equation}
We allow fluctuations as perturbations to the mean-field solution such as $\bar{S}^a\rightarrow{}M^{-1}(\bar{S}^a+\delta\bar{S}^a)$ with $M$ the normalization factor. We would like to examine the dynamics of fluctuations to gain insights into $\tau_{pth}$ since $d\sim{}\delta{}\bar{S}$ in the simplest approximation. We introduce $\bar{\phi}(t)$ and $\bar{\theta}(t)$ such that $(\bar{S}^x,\bar{S}^y,\bar{S}^z)=(\cos\bar{\phi}\sin\bar{\theta},\sin\bar{\phi}\sin\bar{\theta},\cos\bar{\theta}).$ The fluctuations are  now denoted as  $\bar{\phi}(t)\rightarrow\bar{\phi}(t)+\delta\bar{\phi}(t)$ and $\bar{\theta}(t)\rightarrow\bar{\theta}(t)+\delta\bar{\theta}(t).$ Expanding Eq.~\eqref{eq:absLLG} to leading order of $\delta\bar{\phi}$ and $\delta\bar{\theta}$, we obtain~\cite{appendix}
\begin{equation}
	\left(\begin{array}{cc}    
		\partial_t &  \\
		& \partial_t
	\end{array}\right)\left(\begin{array}{c}
		\delta\bar{\phi} \\
		\delta\bar{\theta}
	\end{array}\right)=\frac{\bar{\mathcal{E}}(t)}{\omega}
\left(\begin{array}{c}
		\delta\bar{\phi} \\
		\delta\bar{\theta}
	\end{array}\right),
\end{equation}
where $\bar{\mathcal{E}}$ is a  $\omega$-independent $2\times{}2$ matrix given by the mean-field solution of $\bar{S}^a(t)$.  Therefore, the fluctuations grow  exponentially as $\sim{}e^{t{}\varepsilon/\omega}$ with $\varepsilon$ the eigenvalues of $\bar{\mathcal{E}}$. Since the increase of fluctuations leads to the onset of the prethermal phase, it follows straightforwardly that $\tau_{pth}\propto\omega$. 
The argument should hold for generic frustrated Floquet systems initialized within the g.s. manifold of its zeroth-order Floquet Hamiltonian.  It is attributed to the fact that, for these initial states, the zeroth order term in the equation of motion is negligible. Instead, the LLG equation is governed by the first-order effective Hamiltonian which is linear in $\omega^{-1}$.  Consequently, the significance of  $H^{(1)}_{\rm eff}$ is to induce the dynamical fluctuations, resulting in a prethermalization timescale proportional to the drive frequency.

{\em Prethermal ObD.---} 
Subsequent to $\tau_{pth}$, the system enters a prethermal regime over a thermalization timescales $\tau_{th}$ before eventually heating up to infinite temperature. In accordance with the prethermalization paradigm, the non-equilibrium system can be described by a thermal Gibbs ensemble at a certain temperature $\beta_{pth}$~\cite{Abanin2017,Canovi2016,Weidinger2017}. As our second key result, we find that the prethermal ObD selected phase exhibits a strong dependence on the initial azimuthal angle $\phi_{in}$. 

To characterize the thermal phase and to track the energy absorption, we investigate the evolution of normalized energy $E^{(0)}(t)=H^{(0)}_{\rm eff}(t)/E_0$, where $E_0=-(2J+K)N/3$ is the g.s. energy of $H^{(0)}_{\rm eff}$. %(Analogous notations are introduced for $E^{(1)},...$.)  
As demonstrated in Fig.~\ref{fig:fig1}(c), in the prethermal phase fluctuations are small, and $E^{(0)}(t)$ exhibits a plateau value, persisting for an exponentially long time as $\tau_{th}\sim{}e^{c\omega}$~\cite{Abanin2017CMP,Mallayya2019}. Fixing $\omega$, the manifestation of prethermal ObD can be directly elucidated by the significant dependence of thermalization timescales $\tau_{th}$ on initial azimuthal angles $\phi_{in}$. As $\phi_{in}$ approaches values around $m{}\pi/2$ ($m=0,1,...$, corresponding to $\hat{x}$ and $\hat{y}$ directions), the associated timescales $\tau_{th}$ are much longer compared to those with $\phi_{in}$'s being close to $\pm{}\pi/4$, see Fig.~\ref{fig:fig1}(d). For instance, for $\phi_{in}=\pi/10$, $E^{(0)}(t)$ remains around its plateau value up to $\tau_{th}\sim{}10^6$. In contrast, for $\phi_{in}=\pi/4$, it already deviates from the plateau towards its infinite-temperature value around $\tau_{th}\sim{}10^4$. 

The prethermal ObD also appears in the relationship between prethermal energy $E^{(0)}_{pth}$ and initial azimuthal angles $\phi_{in}$, where $E^{(0)}_{pth}$ is the average of $E^{(0)}(t)$ over a time window in the prethermal regime; see inset of Fig.~\ref{fig:fig1}(d). We find that $E^{(0)}_{pth}$ exhibits a clear $\pi/2$-periodicity on $\phi_{in}$. Note that  $E^{(0)}_{pth}$ is normalized by a negative constant $E_0$. Therefore, its maxima (minima) at $\phi_{in}=0,\pi$ ($\phi_{in}=\pi/4$ modulo $\pi/2$) signify the lowest (highest) points in the landscape of the real energy.  Indeed, the behavior of $E^{(0)}_{pth}$ clearly resembles the free energy of $H^{(0)}_{\rm eff}$ versus $\phi_{in}$ under the thermal ObD mechanism~\cite{appendix,Henley1989}. We emphasize that this prethermal ObD effect vanishes for $K=0$. 

It is well-known that the thermal ObD appears only for finite but small temperatures. A similar effect can also be observed in prethermal ObD, where the inverse drive frequency $1/\omega$ can be thought of as tuning the effective temperature. When $1/\omega$ is zero (zero temperature), the system stops absorbing energy and cannot prethermalize. When $1/\omega$ is sufficiently large (high temperature), the system absorbs energy too quickly to sustain a prethermal regime. Moreover, the oscillation amplitude of $E^{(0)}_{pth}$ w.r.t. $\phi_{in}$ decreases with increasing $\omega$, by noting that $E^{(0)}_{pth}$ at $\phi_{in}=0.2\pi$ increases as  $\omega$ increases and that at $\phi_{in}=0$ slightly changes with $\omega$ (not shown); see the inset of Fig.~\ref{fig:fig1}(c).
%It indicates that in the high-frequency limit, the prethermal ObD is more evident as $\omega$ decreases. This behavior of $E^{(0)}_{pth}$ is qualitatively consistent with low-temperature behavior of free energy in thermal ObD.

\begin{figure}[!t]
	\includegraphics[width=\linewidth]{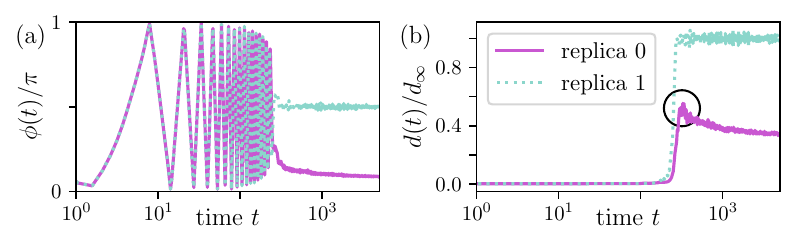}
	\caption{State selection. The dynamics of (a) averaged azimuthal angles $\phi(t)$ and (b) decorrelators $d(t)$ for two different replicas.  Here we used $L=72$, $K=1$, $J^z=0$, $G=10^{-3}$, $\phi_{in}=0.2\pi$, and $\omega=12$.}\label{fig:fig2}
\end{figure}	

\begin{figure}[!t]
	\includegraphics[width=\linewidth]{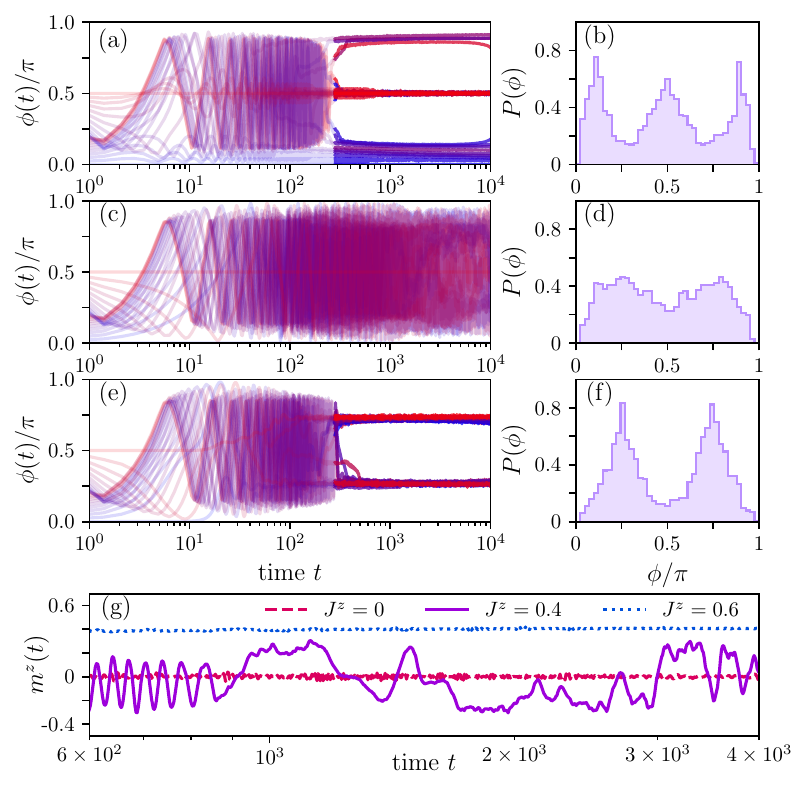}
	\caption{Thermal and prethermal phase transition by $J^z$ in $H^{(0)}_{\rm eff}+H^{(1)}_{\rm eff}$. (a, c, e) Stroboscopic dynamics of averaged azimuthal angles $\phi(t)$ versus initial azimuthal angles $\phi_{in}$ varying from $\phi_{in}=0$ to $\phi_{in}=\pi/4$ (blue to red) for (a) $J^z=0$, (c) $J^z=0.4$, and (e) $J^z=0.8$.  The solid regimes denote the prethermal phases. Here we have used $L = 72$, $K=1$,  $G=10^{-3}$, and $\omega=9$.  b, d, f) Probability distribution $P(\phi)$ for the azimuthal angle obtained by using Monte Carlo simulations with $L=32$, $K=1$, $T=0.08$, $\beta=5.0$, and (b) $J^z=0$, (d) $J^z=0.4$, as well as (f) $J^z=0.8$. (g) The dynamics of out-of-plane magnetization $m^z(t)$ for various $J^z$. The rest model parameters are the same as those in (a, c, e). }\label{fig:fig3}
\end{figure}

One of the crucial aspects of thermal ObD is that it is the entropy that selects the ordered states. Here, in the absence of entropy for individual trajectories, we examine the averaged azimuthal angle,
$\phi(t)=N^{-1}\sum_{r}\cos^{-1}\left(S^{x}_r(t)/\sqrt{1-[S^z_r(t)]^2}\right).$
For different initial angles $\phi_{in}$, we find that $\phi(t)$, after $\tau_{pth}$, reaches several prethermal azimuthal angles such as $\phi(t)\approx{}0$ (modulo $\pi/2$); see Figs.~\ref{fig:fig2}(a) and \ref{fig:fig3}(a). This is consistent with the idea of prethermal ObD, e.g.,  the prethermal angle is selected by the fluctuations generated by the periodic drive.  This is corroborated by the dynamics of the decorrelator $d(t)$ in Fig.~\ref{fig:fig2}(b). After its exponential increase to a relatively large value at $\tau_{pth}$,  $d(t)$ either (i) drops to a relatively smaller plateau value, or (ii) further grows up and stays at a larger plateau value of $d\approx{}1$.  Note that there are four degenerate states selected by the ObD mechanism. 
%It indicates that the system exhibits strong fluctuations around the degenerate manifold, resulting in a large entropy and consequently causing a substantial value of $d(t)$ at $\tau_{pth}$. 
Then, two initial states can end up in the same prethermal state (for instance, both are along the $\hat{x}$ direction) leading to a reduction in the value of $d(t)$. On the other hand, they can occupy two different states (for instance, one along the $\hat{x}$ direction and the other along the $\hat{y}$ direction). Then it can cause a larger plateau value of $d\approx{}\sqrt{1^2+1^2}/d_\infty\approx{}1$.

According to the $Z_4$ symmetry of $H^{(0)}_{\rm eff}$, one would have expected that the prethermal state with $\phi(t)\approx{}0$ (along the $\hat{x}$ direction) and that with $\phi(t)\approx{}\pi/2$ (along the $\hat{y}$ direction) are exactly degenerate.  However, one can find that the prethermal angles $\phi(t)$ deviate slightly from these values; see Fig.~\ref{fig:fig1}(a).  This can be resolved by taking into account the subleading effective Hamiltonian $H^{(1)}_{\rm eff}$~\eqref{eq:H1} breaking $Z_4$ into $Z_2$. We carry out Monte Carlo (MC) simulations for the static Hamiltonian $H^{(0)}_{\rm eff}+H^{(1)}_{\rm eff}$ to sample the thermal ensemble, by which we construct a probability distribution for the azimuthal angle $P(\phi)$. As shown in Fig.~\ref{fig:fig3}(b), the maxima in $P(\phi)$ are consistent with the plateau values in the prethermal regime shown in Fig.~\ref{fig:fig3}(a).  Note that the peak at $\phi=\pi/2$ is broader than the other two peaks, explaining the fact that the fluctuations of $\phi(t)$ around the plateau value of $\pi/2$ are a little bit stronger than those around the other two plateau values. Finally, for comparison, we also perform MC simulations for $H^{(0)}_{\rm eff}$ only, leading to a probability distribution $P(\phi)$ preserving the $Z_4$ symmetry.  

{\em Dynamical phase diagram.--}
The first-order effective Hamiltonian $H^{(1)}_{\rm eff}$ favors noncoplanar order but its contribution to the prethermal energy is small when $J^z=0$ and $K$ is moderately large. As observed above, the dynamics is still governed by prethermal ObD, in which an in-plane FM order is selected. However, increasing $J^z$ leads to a nonvanishing $H_{zx}$ and $H_{zy}$ in Eq.~\eqref{eq:H1} competing with the lowest order $H^{(0)}_{\rm eff}$.

%Strictly speaking, the $O(2)$ accidentally degeneracy of $H^{(0)}_{\rm eff}$ has already been lifted by $H^{(1)}_{\rm eff}$.  Consequently, $H^{(1)}_{\rm eff}$ (as well as higher-order terms in the high-frequency expansion) introduces instabilities to the prethermal ObD mechanism governed by $H^{(0)}_{\rm eff}$ solely. When $J^z$ vanishes and $K=1$ is moderately large, as we have already observed, the non-equilibrium dynamics of $\tau_{th}$, $E^{(0)}_{pth}$, and the selections of $\phi(t)$ can be understood consistently within the framework of ObD mechanism. It indicates that in this limit, 

As a function of increasing $J_z$, we can identify two phases, with a phase transition point at $J^{z,*}\approx0.4$, by monitoring the dynamics of the out-of-plane magnetizations, $m^z(t)=\frac{1}{N}\sum_{r}S^{z}_r(t).$ 
(i) When $J^z<J^{z,*}$ the system clearly manifests prethermal ObD phenomenon, e.g. the prethermal energetics are governed by $H^{(0)}_{\rm eff}$ and the main effect of $H^{(1)}_{\rm eff}$ is to introduce fluctuations as discussed above. In this prethermal ObD phase, $m^z(t)$ vanishes, see Fig.~\ref{fig:fig3}(g). 
(ii) Around the transition point $J^z\approx{}J^{z,*}$, the system is unable to prethermalize, e.g., the timescale of $\tau_{pth}>10^4$ is extremely long; see Fig.~\ref{fig:fig3}(c,g). Neither the azimuthal angles $\phi(t)$ nor the out-of-plane $m^z(t)$ can manifest a well-defined plateau value; instead, they keep displaying strong fluctuations. (iii) For larger $J^z>J^{z,*}$, the [111] phase is stabilized, in which a finite out-of-plane magnetization emerges with a plateau value of $m^z(t)\approx{}\pm{}0.5$. Meanwhile, the azimuthal angle $\phi(t)$ settles to a plateau value shifted towards $\approx{}\pi/4$ (modulo $\pi/2$); see Fig.~\ref{fig:fig3}(e). 

These nonequilibrium dynamics can be understood better with the help of MC simulations for $H^{(0)}_{\rm eff}+H^{(1)}_{\rm eff}$.  For the transition point, the probability distribution for the azimuthal angle $\phi$ (and also the out-of-plane magnetization $m^z$, not shown) has no obvious peaks and valleys, as shown in  Fig.~\ref{fig:fig3}(d). Therefore, the transition point is akin to a critical point with huge thermal entropy, which does not support any orders.  This entropy, manifesting in the non-equilibrium dynamics, prevents the system from rapid prethermalization. On the other hand, in the [111] phase, we find that the probability distribution $P(\phi)$ exhibits two peaks at $\phi=\pi/4$ and $\phi=3\pi/4$, respectively; see Fig.~\ref{fig:fig3}(f). It is consistent with the prethermal FM order observed in Fig.~\ref{fig:fig3}(e), along the [111] direction or its $Z_2$-symmetric counterparts. We note that the maxima of the prethermal energy $E^{(0)}_{pth}$ are still around $\phi=0$ and $\phi=\pi/2$~\cite{appendix}, which are {\em not} the typical prethermal angle of the [111] order. It implies that the prethermal [111] order is not selected by the ObD mechanism of $H^{(0)}_{\rm eff}$, but selected by $H^{(0)}_{\rm eff}+H^{(1)}_{\rm eff}$.

{\em Discussion and conclusion.---}
In addition to known types of `disorder' in equilibrium ObD -- e.g. quenched lattice disorder, thermal and quantum fluctuations -- we showed that fluctuations induced by a periodic drive can select a discrete set of states starting with initial conditions within the accidental degenerate g.s. manifold of the lowest order Floquet Hamiltonian. Focusing on the Floquet XXZ-compass model, we reveal the prethermal ObD through the analysis of the prethermal energy and thermalization time dependence on the initial state direction.  We argue that the concept of prethermal ObD is generic and not sensitive to specific driving protocols. Indeed, we have confirmed similar results for a driven $J_1$-$J_2$ antiferromagnetic XX model~\cite{appendix}.

A remarkable finding is that the prethermal ObD can only be observed after an unusual prethermalization timescale linearly in the drive frequency $\omega$. The reason is that the initial state lies within the accidental degenerate g.s. manifold of the zeroth order effective Floquet Hamiltonian. %A perturbation analysis based on the LLG equation has been developed to support our observation. 
It is then the subleading $\omega-$dependent correction that introduces the fluctuations necessary for the entropic ObD selection, which we confirmed in a fluctuation calculation. As a side effect, we also observe that for large $J^z$ these corrections can trigger a prethermal transition to a different state with out-of-plane magnetization.

For future work, it will be very worthwhile to study other examples of prethermal ObD, e.g., those with an extensive number of accidental symmetries of the g.s. manifold like the antiferromagnetic Heisenberg model on the kagome lattice. In addition, our set-up is a natural platform for investigating the (accidental) symmetry-breaking hierarchy which can be engineered in Floquet systems~\cite{Fu2024}.
Another intriguing possibility is that different choices of the periodic drive can activate different types of fluctuations which could lead to distinct orders in the prethermal plateau. It will also be interesting to establish the general relation between thermal, quantum, and prethermal ObDs. Since entropy plays a pivotal role in achieving prethermal ObD, it will be helpful to develop numerical tools and concepts to quantify entropy production in driven systems. Similarly, it would be worthwhile to think of different experimental platforms for its realization.

In conclusion, we expect that prethermal ObD will provide a versatile playground for intriguing nonequilibrium physics and adds another aspect to one of the paradigms of frustrated systems.

\begin{acknowledgments}
	We thank Marin Bukov, Roderich Moessner and Hongzheng Zhao for helpful discussions and Andrea Pizzi for collaboration on related work. 
	We acknowledge support from the Imperial-TUM flagship partnership, the Deutsche Forschungsgemeinschaft (DFG, German Research Foundation) under Germany's Excellence Strategy--EXC--2111--390814868, DFG grants No. KN1254/1-2, KN1254/2-1, and TRR 360 - 492547816; and from the European Research Council (ERC) under the European Unions Horizon 2020 research and innovation programme (Grant Agreements No. 771537 and No. 851161), the International Centre for Theoretical Sciences (ICTS) for the program "Frustrated Metals and Insulators" (code: ICTS/frumi2022/9), as well as the Munich Quantum Valley, which is supported by the Bavarian state government with funds from the Hightech Agenda Bayern Plus.
\end{acknowledgments}

\bibliography{DN.bib}

\clearpage
\newpage

\begin{widetext}
\begin{center}
\begin{large}
	{\bf Supplementary~Materials~for~``Floquet~Prethermal~Order~by~Disorder''}\\
\end{large}
\end{center}

\begin{center}

{Hui-Ke Jin${}^{1,2}$ and Johannes Knolle${}^{1,2,3}$} \\
{\it $^{1}$Technical University of Munich, TUM School of Natural Sciences, Physics Department, 85748 Garching, Germany} \\
{\it $^{2}$Munich Center for Quantum Science and Technology (MCQST), Schellingstr. 4, 80799 M{\"u}nchen, Germany} \\
{\it $^{3}$Blackett Laboratory, Imperial College London, London SW7 2AZ, United Kingdom}\\

\end{center}

{\color{white} ~}\\

\setcounter{equation}{0}
\setcounter{figure}{0}
\renewcommand{\theequation}{S\arabic{equation}}
\renewcommand{\thefigure}{S\arabic{figure}}
\renewcommand{\thetable}{S\arabic{table}}

In this Supplementary Materials, we provide more information about (i) the stroboscopic equation of motion, (ii) the thermal order-by-disorder effect in $H^{(0)}_{\rm eff}$, (iii) the dynamics for decorrelator $d(t)$, (iv) the perturbation theory based on Landau-Lifshitz-Gilbert equations, (v) prethermal energy for various $J^z$, and (vi) the simulations on the $J_1-J_2$ model. 

\section{Stroboscopic equation of motion for Eq.~(1)}

We consider the dynamical Hamiltonian defined in Eq.~(1) in the main text, which now is re-expressed explicitly as 
\begin{equation*}
H(t) = 
\begin{cases}
-\sum_{r}(J+K)S^x_{r}S^x_{r+\hat{x}}-\sum_{r}JS^x_{r}S^x_{r+\hat{y}} & {\rm~for~} t\in[0,~T/3)\\
-\sum_{r}JS^y_{r}S^y_{r+\hat{y}}-\sum_{r}(J+K)S^y_{r}S^y_{r+\hat{x}} & {\rm~for~} t\in[T/3,~2T/3)\\
-\sum_{r}J^z\left(S^z_{r}S^z_{r+\hat{x}}+S^z_{r}S^z_{r+\hat{y}}\right) & {\rm~for~} t\in[2T/3,~T),
\end{cases}
\label{eq:Ht}
\end{equation*}
As mentioned in the main text, the equation of motion for the above formula is governed by Hamilton’s formalism as 
$\partial_t{}\vec{S}_r=\{\vec{S}_r,H(t)\}.$
The Poisson bracket relation for classical $O(3)$ spins reads
$$\{S^a_{i},S^b_{j}\}=\delta_{i,j}\epsilon^{abc}S^c_i,$$
where $\epsilon^{abc}$ is the fully antisymmetric tensor. Integrating the equation of nation over one total period $T$ analytically, the stroboscopic equation of motion for a spin is $$\vec{S}_r(t+T) = \mathsf{R}_{z}[\theta^z_r(t)]\mathsf{R}_{y}[\theta^y_r(t)]\mathsf{R}_{x}[\theta^x_r(t)]\vec{S}_{r}(t),$$
where $\mathsf{R}_{a}[\theta^a_r]$ ($a=x,y,z$) denotes rotation matrices about the $a$ axis by an angle $\theta^a_r$. Here, the explicit forms of $\theta^a_r$ are expressed as 
\begin{equation*}
\begin{split}
&\theta^x_r = -\frac{T}{3}(J+K)\left(S^x_{r+\hat{x}}+S^x_{r-\hat{x}}\right)-\frac{T}{3}J\left(S^x_{r+\hat{y}}+S^x_{r-\hat{y}}\right),\\
&\theta^y_r = -\frac{T}{3}J\left(S^y_{r+\hat{x}}+S^y_{r-\hat{x}}\right)-\frac{T}{3}(J+K)\left(S^y_{r+\hat{y}}+S^y_{r-\hat{y}}\right),\\
&\theta^z_r = -\frac{T}{3}J^z\left(S^z_{r+\hat{x}}+S^z_{r-\hat{x}}+S^z_{r+\hat{y}}+S^z_{r-\hat{y}}\right).\\    
\end{split}
\end{equation*}
With this integrated equation of motion over long timescales, the classical systems can be simulated efficiently.
%$\theta^a_r= -\frac{T}{3}\sum_{\delta=\pm{}\hat{x},\pm{}\hat{y}}J^a_{\delta{}}S^a_{r+\delta{}}$ (we use the notation of $J^a_{-\delta}\equiv{}J^a_{\delta}$  and $J^z_\delta{}\equiv{}J^z$).

\section{Thermal order by disorder in the 0th-order effective Hamiltonian}

For $K>0$, $H^{(0)}_{\rm eff}$ only preserves the $Z_4$ symmetry. However, the ground state manifold, which corresponds to a ferromagnetic (FM) order, is of $O(2)$. For $J_z<1$, it is obvious that the $z$ component of spins in the ground state manifold should vanish to gain energy. We parameterize the O(3) spins in the ground states as
\begin{equation}
\vec{S}_i=\left(\cos\theta_0, \sin\theta_0, 0\right).
\end{equation}
For a $N=L_x\times{}L_y$ system, the ground state energy is independent of $\theta_0$, which reads
\begin{equation}
E^{(0)}_{\rm eff} = -2JN - KN.
\end{equation}

Now we consider thermal fluctuation within the XY plane, i.e., the spins are still XY spins with 
$$\theta_i\rightarrow{}\theta_0+\delta{}\theta_i.$$
Expand the Hamiltonian around the ground state, we have
\begin{equation}
\begin{split}
	\delta{}H^{(0)}_{\rm eff} =& -\sum_{i,\delta=x,y}J\cos(\delta\theta_i-\delta\theta_{i+\delta})\\
	&-\sum_{i}K\left[\cos(\theta_0+\delta\theta_i)\cos(\theta_0+\delta\theta_{i+x})+\sin(\theta_0+\delta\theta_i)\sin(\theta_0+\delta\theta_{i+y})\right] \\
	=& E_0 + \delta{}E_0 + O^3(\delta\theta_i),
\end{split}
\end{equation}
where 
\begin{equation}
\delta{}E_0 = -\sum_{i}\left[(2J+K)(\delta\theta_i)^2
-(J+K\cos^2\theta_0)\delta\theta_i\delta\theta_{i+x}
-(J+K\sin^2\theta_0)\delta\theta_i\delta\theta_{i+y}
\right].
\end{equation}
Then, we use the Fourier transformations and obtain that 
\begin{equation}
\delta{}E_0 = -\sum_{{\bf q}}\left[(2J+K)
-(J+K\cos^2\theta_0)\cos{}q_x
-(J+K\sin^2\theta_0)\cos{}q_y
\right]\delta\theta_{\bf q}\delta\theta_{-\bf q}.
\end{equation}
Given such a Gaussian form of the low-energy Hamiltonian of $\delta{}H^{(0)}_{\rm eff}$, the corresponding free energy at temperature $1/\beta$ reads $$F\equiv{}-\frac{1}{\beta}\log\left\{\int\mathcal{D}{}[\delta\theta_{\bf q}]\mathcal{D}{}[\delta\theta_{\bf -q}]{}e^{-\beta\delta{}E_0}\right\}.$$ 
We numerically calculate $F$ for given $\theta_0$ and find that the minima of free energy are achieved when 
\begin{equation}
\cos^2\theta_0=0 \textrm{~~or~~}  1.
\end{equation}
Therefore, the $O(2)$ ground state degeneracy is lifted by arbitrary small thermal fluctuations, with FM orders along the $\pm{}\hat{x}$ or ${\pm}{y}$ directions being selected.

\section{Non-equilibrium dynamics for decorrelator}
To probe the signal of chaos and evaluate the relaxation timescale $\tau_{pth}$, we introduce a decorrelator $d(t)$, measuring the 2-norm distance between two initially very close replicas of system, $\vec{S}_r(t)$ and $\vec{S}'_r(t)$, as
\begin{equation}
d(t)=\sqrt{\frac{1}{N}\sum_{r}\left[\vec{S}_r(t)-\vec{S}'_r(t)\right]^2}/d_{\infty}.
\end{equation}
We parameterize the two replicas as 
\begin{equation}
\vec{S}_r=\left(\sin\theta\cos\phi,\sin\theta\sin\phi,\cos\theta\right)~~~{\rm and}~~~
\vec{S}'_r=\left(\sin\theta'\cos\phi',\sin\theta'\sin\phi',\cos\theta'\right),
\end{equation}
where $\theta'_r(t=0)=\theta_r(t=0)+\pi\Delta\epsilon_r$ and $\phi'_r(t=0)=\phi_r(t=0)+2\pi\Delta\varepsilon_r$. Here $\epsilon_r$ and $\varepsilon_r$ are standard normal random numbers and $\Delta{}\ll{}1$ controlling the size of the perturbations.  The spin orientations are completely random (without any correlations) at infinite temperature, and thereby the decorrelator is normalized by an infinite-temperature value of $d_{\infty}=\sqrt{2}$. 

\begin{figure}
\includegraphics[width=0.8\linewidth]{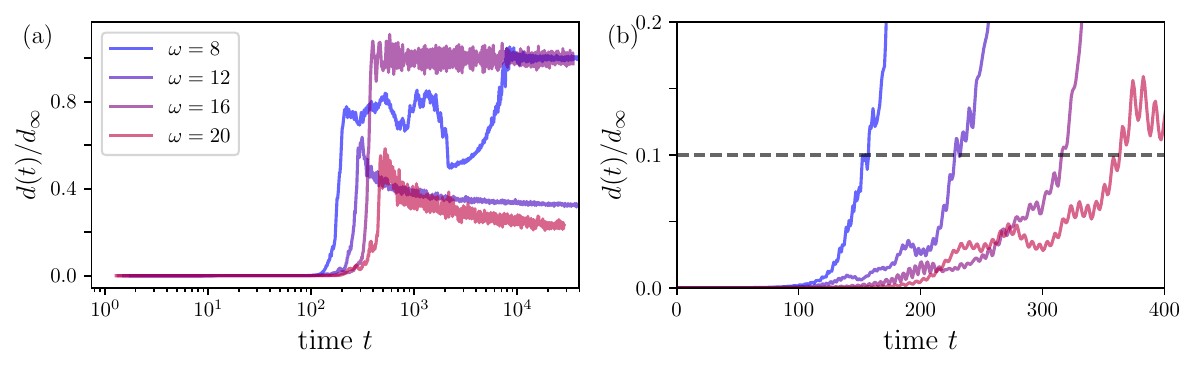}
\caption{(a) The dynamics of the decorrelators for various drive frequency $\omega$. (b) The decorrelator initially grows exponentially. The cross of the dashed line at $0.1$ and $d(t)$ determines the prethermalization timescale $\tau_{pth}$, which exhibits an obvious dependence on $\omega$. }\label{fig:decorrelator}
\end{figure}

In Fig.~\ref{fig:decorrelator} (a),  we show the dynamics of decorrelators for various drive frequencies $\omega$. For short time, the decorrelator grows exponentially from an initial value of $d(t=0)\sim{}\Delta$, as shown in Fig.~\ref{fig:decorrelator}. We quantitatively determine the relaxation timescale $\tau_{pth}$ by the time when the decorrelator $d(t)$ crosses $10\%$ of its infinite temperature value, see Fig.~\ref{fig:decorrelator} (b). Crucially, $\tau_{pth}$, the timescale of the prethermalization, strongly depends on the drive frequency $\omega$. Indeed, we find that $\tau_{pth}\sim{}\omega$, see the figures in the main text, which is unusual in general Floquet heating profiles. And later we will resolve this linear dependence in $\omega$ by a perturbation theory.

%In addition to the observation of $\tau_{pth}$, we can also unveil the prethermal order-by-disorder phenomenon by studying $d(t)$ in the prethermal regime. At the beginning of the prethermal phase, $d(t)$ increases rapidly to a finite value, and then it will (i) drop to a relatively smaller plateau value or (iii) further go up and strongly fluctuate around a larger plateau value of $d\approx{}1$. This behavior is a direct consequence of state selections driven by the prethermal order-by-disorder phenomenon. Note that there are four-fold degenerate states in the prethermal regime, corresponding to the $Z_4$ rotational symmetry.  the system exhibits strong fluctuations between degenerate states, resulting in a large thermal entropy and consequently causing a substantial value of $d(t)$ around the transition time $\tau_{pth}$. After $\tau_{pth}$, one of the degenerate states will be selected. For two replicas with close initial conditions, on the one hand, they can be selected to stay in the same state (for instance, both are along the $\hat{x}$ direction). This can lead to a reduction in the value of $d(t)$, reaching a smaller plateau value, as the system stops fluctuating between degenerate states. On the other hand, they can be chosen to occupy two different states (for instance, one along the $\hat{x}$ direction and the other along the $\hat{y}$ direction). Then it can cause a larger plateau value of $d\approx{}\sqrt{1^2+1^2}/d_\infty\approx{}1$.

\section{Landau-Lifshitz-Gilbert equations for the effective Hamiltonian at $J^z=0$}

We focus on the non-equilibrium dynamics for the Hamiltonian defined in Eq.(1) with $J^z=0$ (see main text). The effective Hamiltonian is simplified as 
\begin{equation}
H_{\rm eff}=\frac{1}{3}\left(H_x+H_y\right)-\frac{\pi}{9\omega}H_{yx} +O(\omega^{-2}).
\end{equation}
The equation of motion for a spin at site $r$ reads
\begin{equation}\label{eq:LLG}
\begin{split}
	\frac{\partial{}S^a_{r}(t)}{\partial t}=&\{S^a_r(t), H_{\rm eff}\}\equiv{}F^{(0)}_{a}(r,t)+F^{(1)}_{a}(r,t) +O(\omega^{-2}),
\end{split}
\end{equation}
where $F^{(n)}_{a}(r)$ corresponds to the $n$-th order effective Hamiltonian $H^{(n)}_{\rm eff}$. This differential equation in Eq.~\eqref{eq:LLG} is a generalized-version Landau–Lifshitz–Gilbert (LLG) equation.
For the zero-th order, we find that 
\begin{equation}
\begin{split}
	F^{(0)}_{x}(r,t)=&-\frac{1}{3}\left[(J+K)\left(S^z_rS^y_{r+\hat{y}}+S^z_rS^y_{r-\hat{y}}\right)+J\left(S^z_rS^y_{r+\hat{x}}+S^z_rS^y_{r-\hat{x}}\right)\right],\\
	F^{(0)}_{y}(r,t)=&+\frac{1}{3}\left[(J+K)\left(S^z_rS^x_{r+\hat{x}}+S^z_rS^x_{r-\hat{x}}\right)+J\left(S^z_rS^x_{r+\hat{y}}+S^z_rS^x_{r-\hat{y}}\right)\right],\\
	F^{(0)}_{z}(r,t)=&-\frac{1}{3}\left[(J+K)\left(S^y_rS^x_{r+\hat{x}}+S^y_rS^x_{r-\hat{x}}\right)+J\left(S^y_rS^x_{r+\hat{y}}+S^y_rS^x_{r-\hat{y}}\right)\right]+\\
	&+\frac{1}{3}\left[(J+K)\left(S^x_rS^y_{r+\hat{y}}+S^x_rS^y_{r-\hat{y}}\right)+J\left(S^x_rS^y_{r+\hat{x}}+S^x_rS^y_{r-\hat{x}}\right)\right].
\end{split}
\end{equation}
And for the first-order we find that,
\begin{equation}
\begin{split}
	F^{(1)}_x(r)=&\frac{\pi}{9\omega}\sum_{\delta=\pm{}\hat{x},\pm{}\hat{y}}\sum_{\delta'\neq{}\delta}J^y_{\delta}J^x_{\delta'}\left(S^z_{r}S^x_{r+\delta-\delta'}S^z_{r-\delta}-S^y_{r+\delta}S^x_{r+\delta'}S^y_{r}\right)+\\
	&\frac{\pi}{9\omega}\sum_{\delta=\hat{x},\hat{y}}J^x_{\delta}J^y_{\delta}\left(S^z_rS^x_rS^z_{r+\delta}+S^z_rS^x_rS^z_{r-\delta}-S^y_{r-\delta}S^x_{r-\delta}S^y_{r}-S^y_{r+\delta}S^x_{r+\delta}S^y_{r}\right),\\
	F^{(1)}_y(r)=&\frac{\pi}{9\omega}\sum_{\delta=\pm{}\hat{x},\pm{}\hat{y}}\sum_{\delta'\neq{}\delta}J^y_{\delta}J^x_{\delta'}\left(S^y_{r+\delta}S^x_{r+\delta'}S^x_{r}-S^y_{r+\delta-\delta'}S^z_{r-\delta'}S^z_{r}\right)+\\
	&\frac{\pi}{9\omega}\sum_{\delta=\hat{x},\hat{y}}J^x_{\delta}J^y_{\delta}\left(S^y_{r-\delta}S^x_{r-\delta}S^x_{r}+S^y_{r+\delta}S^x_{r+\delta}S^x_{r}-S^y_{r}S^z_{r}S^z_{r+\delta}-S^y_{r}S^z_{r}S^z_{r-\delta}\right),\\
	F^{(1)}_{z}(r)=&\frac{\pi}{9\omega}\sum_{\delta=\pm{}\hat{x},\pm{}\hat{y}}\sum_{\delta'\neq{}\delta}J^y_{\delta}J^x_{\delta'}\left(S^y_{r}S^z_{r-\delta}S^y_{r+\delta'-\delta}-S^x_{r}S^z_{r-\delta'}S^x_{r+\delta-\delta'}\right)+\\
	&\frac{\pi}{9\omega}\sum_{\delta=\hat{x},\hat{y}}J^x_{\delta}J^y_{\delta}\left(S^y_rS^y_r-S^x_rS^x_r\right)\left(S^z_{r+\delta}+S^z_{r-\delta}\right).
\end{split}
\end{equation}
Note that, for simplicity, the time index $t$ for spins has been omitted.

\begin{figure}
\includegraphics[width=0.75\linewidth]{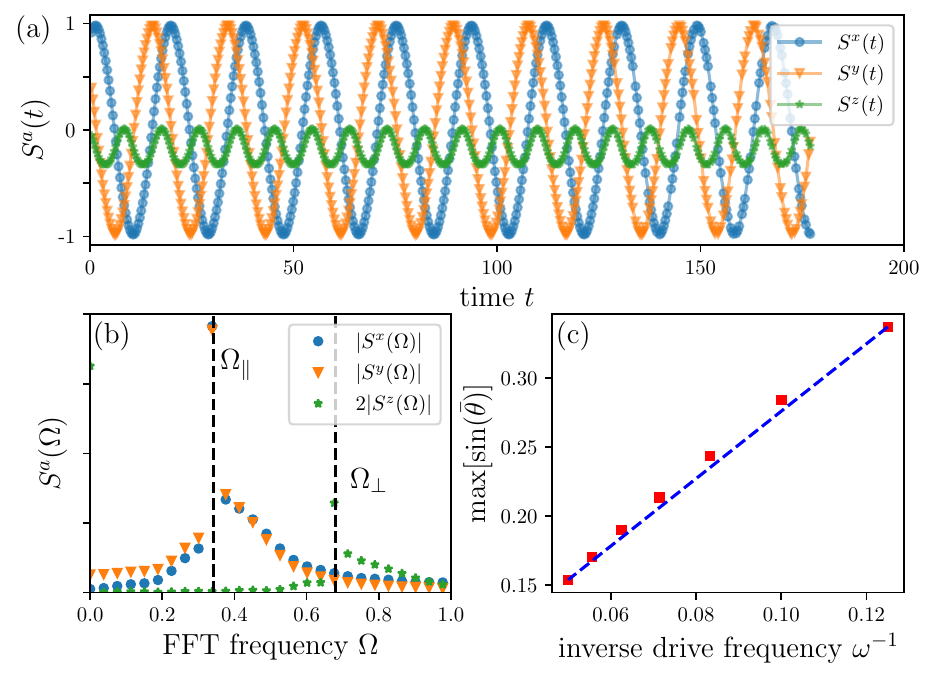}
\caption{(a) The short-time dynamics for the averaged spin magnetizations $S^a(t)$ ($a=x,y,z$). We emphasize that the variance of spin magnetizations $\Delta{}S^a(t)$ is always around $\sim{}10^{-4}$ for time $t<200$. Therefore, the systems can be treated as a single spin. Here the model parameters are $J=K=1$, $J^z=0$, linear lattice size $L=72$, drive frequency $\omega=18$, initial Gaussian noise $G=10^{-4}$, and initial azimuthal angle $\phi_{in}=0.2\pi$. (b) The Fourier transformation of the short-time $S^a(t)$ is shown in panel (a). We find that $S^a(\Omega)$  exhibits a peak around $\Omega_{\perp}$ for $a=x,y$ and that around $\Omega_{\perp}$ for $a=z$. (c) The out-of-plane magnetization $\sin\bar\theta$ exhibits a clear linear dependence of inverse drive frequency $\omega^{-1}$. }\label{fig:figc1}
\end{figure}

We initialize the evolution with a FM ordered state perturbed by tiny Gaussian fluctuations. Consequently, at least for the short dynamics, the system can be well described as a mean-field ansatz as $S^a_r(t)\rightarrow\bar{S}^a(t)$, namely, all spins rotate simultaneously as a single spin, see Fig.~\ref{fig:figc1}(a). With this treatment, we can rewrite Eq.~\eqref{eq:LLG}
\begin{equation}
\begin{split}
	\frac{\partial{}\bar{S}^x(t)}{\partial{}t}&=-\frac{2(K+2J)}{3}\bar{S}^z\bar{S}^y+\frac{4\pi(K+2J)^2}{9\omega}(\bar{S}^z\bar{S}^z-\bar{S}^y\bar{S}^y)\bar{S}^x,\\
	\frac{\partial{}\bar{S}^y(t)}{\partial{}t}&=+\frac{2(K+2J)}{3}\bar{S}^z\bar{S}^x-\frac{4\pi(K+2J)^2}{9\omega}(\bar{S}^z\bar{S}^z-\bar{S}^x\bar{S}^x)\bar{S}^y,\\
	\frac{\partial{}\bar{S}^z(t)}{\partial{}t}&=\frac{4\pi(K+2J)^2}{9\omega}(\bar{S}^y\bar{S}^y-\bar{S}^x\bar{S}^x)\bar{S}^z.\label{eq:SLLG}
\end{split}
\end{equation}
Note that the leading order for $\bar{S}^z$ is given by $F^{(1)}_z$ since  $F^{(0)}_z$ always vanishes with such a mean-field  treatment. Nevertheless, $F^{(0)}_x$ and $F^{(0)}_y$ can be finite. The mean-field equation of motion can be further simplified by introducing $\bar{\phi}(t)$ and $\bar{\theta}(t)$ to parameterize the exact mean-field solution as
\begin{equation}\label{eq:mfsolution}
\left[\bar{S}^x(t),\bar{S}^y(t),\bar{S}^z(t)\right]=\left[\cos\bar{\phi}(t)\sin\bar{\theta}(t),\sin\bar{\phi}(t)\sin\bar{\theta}(t),\cos\bar{\theta}(t)\right].
\end{equation}
Then Eqs.~\eqref{eq:SLLG} are reduced into two independent equations as
\begin{equation}
\begin{split}
	&\frac{\partial{}\bar{\phi}}{\partial{}t}=\frac{2(K+2J)}{3}\cos\bar{\theta}-\frac{\pi(K+2J)^2}{9\omega}\left(1+3\cos2\bar{\theta}\right)\sin2\bar{\phi},\\
	&\frac{\partial{}\bar{\theta}}{\partial{}t}=\frac{\pi(K+2J)^2}{9\omega}\left(1+3\cos2\bar{\theta}\right)\sin2\bar{\theta}\cos2\bar{\phi},\label{eq:LLG3}
\end{split}
\end{equation}
When $\omega\rightarrow\infty$, namely, $F^{(1)}_a(r,t)\rightarrow{}0$, one can find that $\bar\theta(t)=\pi/2$ can be the exact solutions for Eq.~\eqref{eq:LLG3}. As shown in Figs.~\ref{fig:figc1}(a) and (b), $\sin\bar{\theta}$ and $\cos\bar{\phi}$ exhibit  harmonic behaviors with frequencies $\Omega_{\perp}$ and $\Omega_{\parallel}$, respectively. We numerically find that $\Omega_{\perp}=2\Omega_{\parallel}$ and $\Omega_{\perp}\propto{}1/\omega$, with $\omega$ the drive frequency. 
%Moreover, we find that $J^x_{\delta}=J^y_{\delta}=J$ and thereby arbitrary initial $\phi(t)$ leads to the same $F^{(0)}_{a}$ ($a=x,y,z$). It indicates the vanishing of prethermal order-by-disorder effect at $K=0$.

We now  allow weak fluctuations around the mean-field solution as  
$$\bar{\phi}(t)\rightarrow\bar{\phi}(t)+\delta\phi(t),~~~\bar{\theta}(t)\rightarrow\bar{\theta}(t)+\delta\theta(t),$$
where $\delta{}\phi(t),\delta{}\theta(t)\ll{}1$ can be treated as small perturbations.   Note that here, for simplicity, we have assumed that the fluctuations are independent of spacial dimension, namely, lattice site $r$.  Since $\bar{S^z}$ is relatively small, $\delta\theta_r$ and $\delta\phi_r$ can be approximately considered as the out-of-plane and in-plane fluctuations, respectively.
Then, by expanding Eq.~\eqref{eq:LLG} in the leading order of perturbations, we find two independent equations of motion as
\begin{equation}
\left(\begin{array}{cc}
	\partial_t &  \\
	& \partial_t
\end{array}\right)\left(\begin{array}{c}
	\delta\phi \\
	\delta\theta
\end{array}\right)=\frac{\bar{\mathcal{E}}}{\omega}\left(\begin{array}{c}
	\delta\phi \\
	\delta\theta
\end{array}\right) \equiv
\frac{1}{\omega}\left(\begin{array}{cc}
	B_{11} & B_{12}\\
	B_{21} & B_{22}
\end{array}\right)\left(\begin{array}{c}
	\delta\phi \\
	\delta\theta
\end{array}\right),\label{eq:dllg1}
\end{equation}
where 
\begin{equation*}
\begin{split}
&B_{11}=-\frac{2\pi(K+2J)^2}{9}\left(1+3\cos2\bar{\theta}\right)\cos2\bar{\phi},\\ 
 &B_{12}=-\frac{2(K+2J)}{3}\omega\sin\bar{\theta}+\frac{6\pi(K+2J)^2}{9}\sin2\bar{\theta}\sin2\bar{\phi},\\ 
 &B_{21}=-\frac{2\pi(K+2J)^2}{9}\left(1+3\cos2\bar{\theta}\right)\sin2\bar\theta\sin2\bar{\phi},\\ &B_{22}=\frac{2\pi(K+2J)^2}{9}\left(\cos2\bar{\theta}+3\cos4\bar{\theta}\right)\cos2\bar{\phi}.
\end{split}
\end{equation*}
Note that $B_{ab}$ are time-dependent factors. It seems to be that the first term in $B_{12}$ is proportional to $\omega$. However, it is evident that $\sin\bar{\theta}$, characterizing the out-of-plane magnetization, exhibits a linear dependence of $\omega^{-1}$; see Fig.~\ref{fig:figc1}(c). Overall, we numerically confirm that implicitly, the leading order of $\omega\sin\bar{\theta}$ and thereby $B_{12}$ is $\omega^{0}$ rather than $\omega$.

By taking the ansatz of $\delta{}\theta\sim{}e^{t{}E_1}$ and $\delta{}\phi\sim{}e^{t{}E_2}$, one can diagonalize Eq.~\eqref{eq:dllg1} and obtain 
\begin{equation}
	E_1\propto\frac{1}{\omega} {\rm~~and~~} E_2\propto\frac{1}{\omega}.
\end{equation}
Since the increase of fluctuations leads to the beginning of prethermalization, we have 
$$\tau_{pth}\sim{}(E_1)^{-1}\sim{}\omega.$$

\section{Prethermal energy for various $J^z$}
\begin{figure}[!t]
\centering
\includegraphics[width=0.8\linewidth]{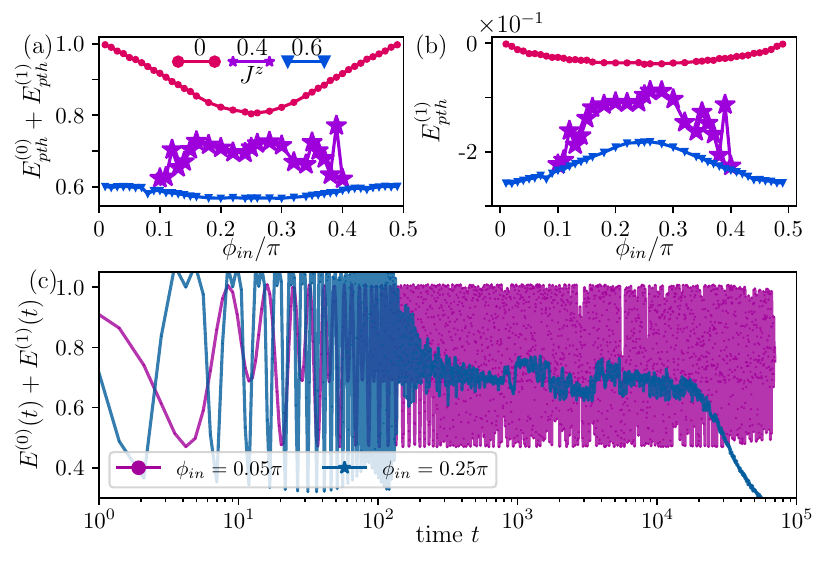}
\caption{(a, b) The prethermal energies (a) $E^{(0)}_{pth}+E^{(1)}_{pth}$ and (b) $E^{(1)}_{pth}$ as functions of initial azimuthal angles $\phi_{in}$ in the prethermal ObD phase ($J^z=0$), around transition point ($J^z=0.4$), and in the dynamical [111] phase ($J^z=0.6$), respectively. (c) The dynamics of $E^{(0)}$ for $J^z=0.4$. Here we use $L = 72$, $K=1$,  $G=10^{-3}$, and $\omega=9.0$.}
\label{fig:figc3}
\end{figure}

In this section, we small more information about the zero-order prethermal energy $E^{(0)}_{pth}$ and also the first-order prethermal energy $E^{(1))}_{pth}$ for different $J^z$. Here $E^{(1))}_{pth}$ is an average of 
$E^{(1))}(t) \equiv{}H^{(1)}_{\rm eff}(t)/E_0$ over a time window in the prethermal regime.

The short-time dynamics of the system are mainly governed by $H^{(0)}_{\rm eff}$ supporting ObD phenomena. Consequently, apart from the prethermal ObD phase, the dynamical [111] phase also manifests, to varying degrees, a dependence on the initial angle $\phi_{in}$. As shown in Fig.~\ref{fig:figc3}(a), the prethermal energy $E^{(0)+(1)}_{pth}\equiv{}E^{(0)}_{pth}+E^{(1)}_{pth}$ in the [111] phase ($J^z=0.6$) demonstrates a $\pi/2$ periodicity, akin to the prethermal ObD phase. However, it does exhibit a reduced oscillation amplitude, indicating now the effect of prethermal ObD is less significant.

The dynamical phase transition can be characterized by the first-order prethermal energy $E^{(1)}_{pth}$, as shown in Fig.~\ref{fig:figc3}(b). In the prethermal ObD phase ($J^z=0$), we find that $E^{(1)}_{pth}$ is at least two orders of magnitude smaller than $E^{(0)}_{pth}$. More importantly, it exhibit a similar $\pi/2$ periodicity with respect to $\phi_{in}$, when compared to those of $E^{(0)}_{pth}$. Therefore, we can conclude that in the prethermal ObD phase, $H^{(1)}_{\rm eff}$ is a perturbative correction to $H^{(0)}_{pth}$. On the contrary, in the dynamical [111] phase ($J^z=0.6$), $E^{(1)}_{pth}$ is only one order of magnitude smaller than $E^{(0)}_{pth}$. Moreover, its $\pi/2$ periodicity is in the opposite phase compared to that of $E^{(0)}_{pth}$. This indicates that $H^{(1)}_{pth}$ is more of a competitor rather than a mere correction to  $H^{(0)}_{pth}$. Around the transition point, as previously mentioned, the system is hard to prethermalize for $\phi_{in}$ being proximity to $m\pi/2$ ($m=0,1,2,...$), as demonstrated in Fig.~\ref{fig:figc3}(c). It is reasonable to anticipate that at the exact transition point, inadequate prethermalization can appear for any value of $\phi_{in}$. For the $\phi_{in}$ for which the system achieve prethermalization, we find that $E^{(1)}_{pth}$ roughly exhibit a $\phi_{in}$-independent behavior. It indicates that due to the prolonged prethermalization time, the information given by initial conditions becomes lost.

\section{Prethermal order-by-disorder in the square-lattice $J_1$-$J_2$ XX model}

We introduce a periodically driven model  on the square lattice. The spins are governed by the following periodic binary Hamiltonian at frequency $\omega=2\pi/T$,
\begin{equation}
	H^{\rm xx}(t)= 
	\begin{cases}
		&J_1\sum_{\langle{}ij\rangle_1}S^x_iS^x_j+J_2\sum_{\langle{}ij\rangle_2}S^x_iS^x_j\quad {\rm~~for~~} t\in[0,~T/2)\\
		&J_1\sum_{\langle{}ij\rangle_1}S^y_iS^y_j+J_2\sum_{\langle{}ij\rangle_2}S^y_iS^y_j\quad {\rm~~for~~} t\in[T/2,~T),
	\end{cases}\label{eq:Hxx}
\end{equation}
where $J_1>0$ and $J_2\geq{}0$ are antiferromagnetic (AFM) XX exchanges, and $\langle{}ij\rangle_1$ as well as $\langle{}ij\rangle_2$ denote the 1st and 2nd nearest-neighbor bonds, respectively.  The effective Hamiltonian for Eq.~\eqref{eq:Hxx} in the leading order is just a $J_1$-$J_2$ XX model with  in-plane $O(2)$ rotational symmetry, which reads
\begin{equation}
	H^{\rm xx}_{\rm eff}=\frac{J_1}{2}\sum_{\langle{}ij\rangle_1}\left(S^x_iS^x_j+S^y_iS^y_j\right)+\frac{J_2}{2}\sum_{\langle{}ij\rangle_2}\left(S^x_iS^x_j+S^y_iS^y_j\right). \label{eq:Hxxeff}
\end{equation}
The ground-state phase diagram for this model has been investigated: (i) If $J_2/J_1<0.5$, the ground state is the well-known AFM N\'eel order. (ii) When $J_2/J_1\geq{}0.5$, the system breaks up into two square sublattices, say $a$ and $b$. In $a$ ($b$) sublattice, the spins form a  N\'eel order, in which the direction of the N\'eel order can be characterized by a angle $\vartheta_a$ ($\vartheta_b)$. The ground states, up to a global $O(2)$ rotation, can be labeled by a reference angle $\vartheta\equiv\vartheta_a-\vartheta_b$.  Here we name this kind of ground state as a two-sublattice state with a reference angle $\vartheta$.

\begin{figure}[!tp]
	\includegraphics[width=0.9\linewidth]{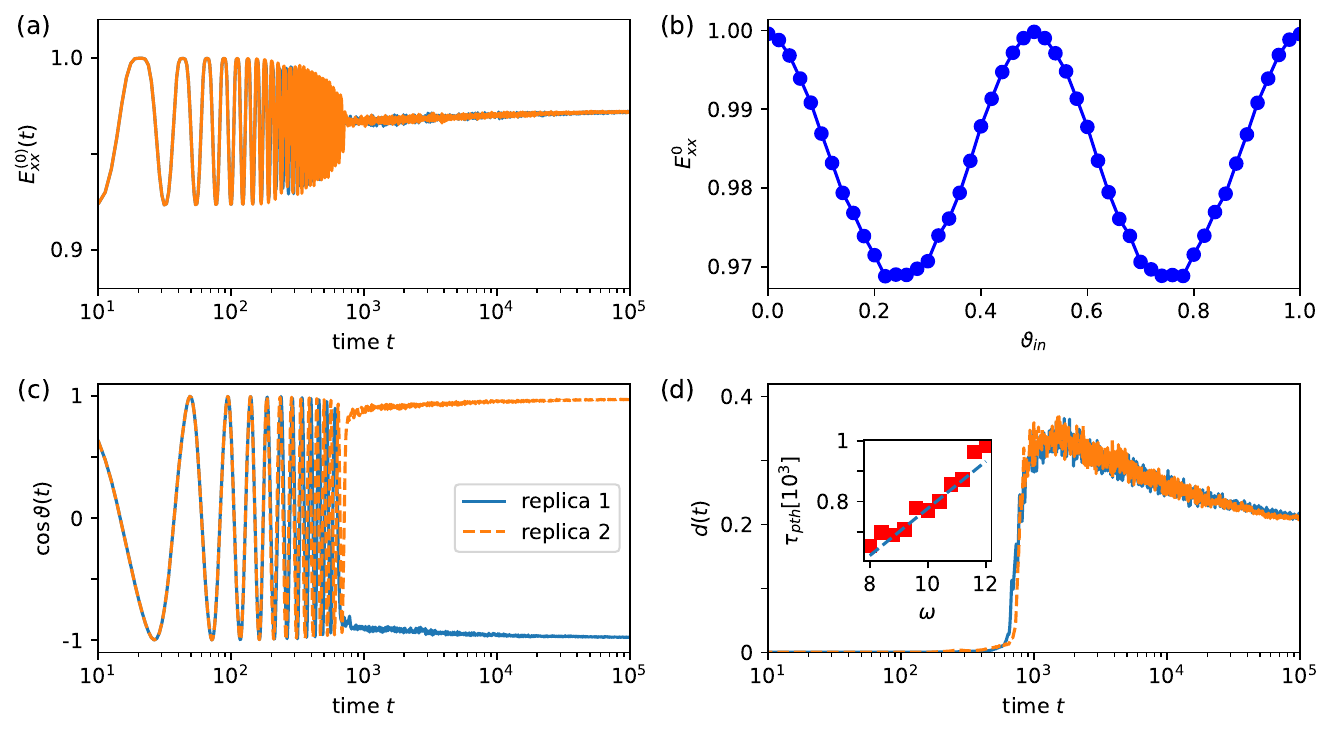}
	\caption{The prethermal order-by-disorder effect in the $J_1$-$J_2$ XX model.  (a) The dynamics for the zero-th energy, $E^{0}_{xx}(t)$, normalized by the g.s. energy of the Hamiltonian~\eqref{eq:Hxxeff}.  (b) The corresponding plateau value of $E^{0}_{xx}(t)$ as a function of the initial reference angles $\vartheta_{in}$.	(c)  The reference angles between $a$ and $b$ sublattices in the prethermal regime will be selected as $\vartheta=0,\pi$. (d) The decorrelator increases rapidly around the prethermal $\tau_{pth}$, and decreases sightly at the onset of the prethermal regime.  Inset: the prethermal timescale as a function of drive frequency $\omega$. We take that the AFM XX exchanges $J_1=J_2=1$, drive frequency $\omega=10$, initial Gaussian notice $G=0.01$, and linear lattice size $L=72$. For panels (a), (c), and (d), the initial reference angle is fixed as $\vartheta_{in}=0.22\pi$. }\label{fig:j1j2}
\end{figure}

Because the ground-state energy for Eq.~\eqref{eq:Hxxeff}, $E^{xx}_0=-N{}J_2$, is independent o $\vartheta$, there is an accidentally degenerate $O(2)$ ground-state manifold. It indicates a thermal order-by-disorder effect at finite temperature, and it has been point out in Ref.~\cite{Henley1989} that the thermal fluctuations will select a collinear two-sublattice state with $\vartheta=0,\pi$ ($\cos\vartheta=\pm{}1$). 

Here, we evolve this system according to the dynamical Hamiltonian in Eq.~\eqref{eq:Hxx}. The initial state is prepared to be a two-sublattice state with initial reference angle $\vartheta_{in}$. 	
In-plane Gaussian noise, parameterized by $G$, is incorporated into the initial state. (see main text for the details). The main observables of interest are the normalized zeroth order energy $E^{(0)}_{xx}(t)$, the reference angles $\cos\vartheta(t)$, and the decorrelator $d(t)$

The main results are summarized in Fig.~\ref{fig:j1j2}. Similar to the dynamical XXZ-compass model discussed in the main text, the plateau value of  $E^{(0)}_{xx}(t)$ also demonstrates a $\pi/2$-periodicity on the initial reference angles $\theta_{in}$, see Fig.~\ref{fig:j1j2}(b). This provides compelling evidence for the prethermal order-by-disorder effect. As shown in Fig.~\ref{fig:j1j2}(c), the periodic drive selects a collinear prethermal state with $\vartheta=0,\pi$, akin to the thermal order-by-disorder mechanism.  We observe a rapid increase in the decorrelator around the prethermal timescale $\tau_{pth}$, followed by a slight decrease at the onset of the prethermal regime. Mover, the inset of Fig.~\ref{fig:j1j2} (d) illustrates that the prethermalization timescale $\tau_{pth}$ exhibits a linear dependence on the drive frequency $\omega$. 

In summary, we demonstrate that the prethermal order-by-disorder effect can emerge as a general non-equilibrium phase of matter, prevalent in diverse periodically driven systems.

\end{widetext}

\end{document}